\documentstyle[twocolumn,psfig]{mn}

\newif\ifAMStwofonts
\AMStwofontstrue
\def\kms{\rm{km s^{-1}}}

\def\calH{{$\cal H$}\,}
\def\gsim{~\rlap{$>$}{\lower 1.0ex\hbox{$\sim$}}}
\def\dhi{n_{_{\rm HI}}}

\def\dhhi{\hat n_{_{\rm HI}}}

\def\nhi{\dhi}

\def\kms{\rm{km s^{-1}}}
\def\Mpc{h^{-1}{\rm Mpc}}

\def\ltsim{\lower.5ex\hbox{$\; \buildrel < \over \sim \;$}}
\def\gtsim{\lower.5ex\hbox{$\; \buildrel > \over \sim \;$}}
\def\ltsim{\lower.5ex\hbox{$\; \buildrel < \over \sim \;$}}
\def\gtsim{\lower.5ex\hbox{$\; \buildrel > \over \sim \;$}}

\def\vx{{\bf x}}

\def\kms{\mbox{km\,s$^{-1}$}}

\def\dd{\,{\rm d}}

\newcommand{\op}{Ly$\alpha$\ }
\newcommand{\hi}{\mbox{H{\scriptsize I}}}

\begin{document}
\title[]{ The \op\ forest in a truncated  hierarchical structure
formation}
\author[Nusser \& Lahav ] {Adi Nusser$^1$ and Ofer Lahav$^{2,3}$\\
$^1$ Physics Department, The Technion-Israel Institute of Technology, Haifa, Israel \\
$^2$ Institute of Astronomy, Madingley Rd., CB3 0HA, Cambridge, UK \\
$^3$ Racah Institute of Physics, The Hebrew University, Jerusalem 91904, Israel\\
}

\maketitle

\begin{abstract}
 The \op\ forest
provides  important constraints on the smoothness
of the universe on large scales.
We calculate the flux distribution along line-of-sight 
to  quasars in a universe made of randomly distributed clumps,
each of them  with a Rayleigh-L`evi fractal structure.
We consider
the  probability distribution function
of the normalised flux in the line-of-sight to quasars.
We show that the truncated clustering hierarchy model 
shows far too many voids along the 
line-of-sight to  quasars compared with the 
observed flux distribution
and the distribution in a Cold Dark Matter model.
This  supports the common view that 
on large scales the universe is homogeneous, rather than  fractal-like.
  
\end{abstract}
\begin{keywords}
cosmology: theory, observation , dark matter, large-scale structure
of the Universe --- intergalactic medium --- quasars: absorption lines
\end{keywords}

\section{Introduction} 

The underlying paradigm in cosmology is the Cosmological
Principle of isotropy and homogeneity (e.g. Peebles 1993).  This
principle was adopted at the time that observational data on the large
scale structure in the universe were not available.  At present it is
well established that the distribution of galaxies and mass
are clumpy on scales smaller than tens of Mpcs.
It is important therefore to quantify the gradual transition from
clumpiness to homogeneity.  This transition can be phrased in terms of
the fractal dimension of the galaxy (or mass) distributions. 
The fractal dimension, $D$,  is defined as 
\begin{equation}
N(<R)\sim R^{D} \, ,
\label{frac:def}
\end{equation}
where $N$ is the mean number of objects within a sphere of radius $R$ centered
on a randomly selected object.
 On
scales $< 10 \Mpc$ the fractal dimension of the galaxy distribution is
$D = 1.2$, but the fluctuations in the X-ray Background and in the
Cosmic Microwave Background on scales larger than $300 \Mpc$ are
consistent with $D=3$ to very high precision (e.g. Peebles 1980; 
Wu, Lahav \& Rees 1999)
in agreement with the Cosmological Principle.  On the other
hand, there are claims (e.g. Pietronero, Montuori \& Sylos-Labini 1997) 
that the
distribution of galaxies is characterized by a fractal with $D=2\pm 0.2$
all the way to
scales of $\sim 500 \Mpc$.  Because of the significance of this issue, it is
important to investigate other independent probes of the matter
distribution in the universe.  Here we attempt to set constraints on
the smoothness on large scales from the distribution of 
the normalized flux in the 
\op\ forest
in the lines-of-sight to distant quasars.  As a probe to the matter
distribution the \op\ forest has a number of advantages over galaxies.  The
forest reflects the neutral hydrogen (\hi ) distribution and therefore
is likely to be a more direct  trace of the mass distribution 
than galaxies are,
especially in low density regions.  Unlike galaxy surveys which are
limited to the low redshift universe, the forest spans a large
redshift interval, typically $2.2 < z < 4$, corresponding 
to comoving interval of $\sim 600 \Mpc$.
Also, observations of the
forest are not contaminated by complex selection effects such as those
inherent in galaxy surveys.  It has been suggested qualitatively by
Davis (1997) that the absence of big voids in the distribution of \op\
absorbers is inconsistent with the fractal model (see also, Wu et
al. 1999). Furthermore, all lines-of-sight towards quasars look
statistically similar.  
Here we predict the distribution of the flux  in \op\ observations in a specific
truncated fractal-like model. We find that indeed in this model there are
too many voids compared with the observations and conventional models
for structure formation.

The outline of the paper is as follows. In section 2 we describe the
truncated clustering hierarchy (TCH) model for the distribution of dark
matter. In section 3 we briefly review the physics of the \op\
and describe how to obtain \hi\ density field from the dark  matter 
 distribution.
In section 4 we show the results for the probability distribution
function of the flux. In section 5 we conclude with a
summary and a general discussion.

\section{the dark matter distribution}

We consider a  dark matter  model that is 
based on a fractal-like distribution of points. 
There are many candidates for this type of models.
Here we will focus on one of the most
studied of these models, namely the Rayleigh-L'evy (RL) walk (Mandelbrot
1977,
Peebles 1980, 1993). As we
will see later, our analysis relies on rather general fractal
properties, so we do not expect our conclusions to depend on the
details of the fractal model.  Our main conclusion is that a fractal
distribution of matter with $D<2$ is inevitably characterized by large voids,
which is inconsistent with observations of the flux in the
\op\ forest.  In practice  we will use the
TCH model in which the distribution
of matter is made of a finite number of Rayleigh-L'evy clumps, 
each  with a
finite number of steps (points). 
This model obviously does not have as many big voids as the
pure fractal distribution. Therefore, if we demonstrate that the TCH
already contains far too many large voids to be 
compatible with \op\ data then
we can safely rule out a pure fractal distribution on the same basis.

In the TCH model, the  
dark matter  distribution in a large volume is represented by a
finite number of points. The points are distributed in
$n_c$ clumps each made of $n_s$ steps (points) so  
 that the total number of
points in the volume is $n_c n_s$.   
The clumps are randomly placed inside the volume and the distribution of
points in each clump is generated from a Rayleigh-L'evy  random
walk (e.g. Peebles 1980) in which the
individual steps have random directions, and random lengths drawn
according to the following cumulative probability function
\begin{eqnarray}
P(>l)&=& \left(\frac{l}{l_0}\right)^{-\alpha} \quad {\mathrm for} \quad
l \ge l_0\\ \nonumber
&&1 \quad {\mathrm otherwise} \\ \nonumber
\end{eqnarray}
For $\alpha> 2 $ the variance of $l$ is finite. Therefore, by the central limit
theorem, the displacement of a point after sufficiently large 
number of steps has
a Gaussian distribution. 
On the other hand, for $ \alpha \le 2$ the variance  is infinite
and the clump follows a {\em truncated fractal} structure where
a pure fractal with dimension $D=\alpha$ is 
obtained  only in the limit of an infinite number of steps.  
In figure
\ref{fig:chain} we show a projection in the plane of a realization of
a three dimensional single RL clump generated using 160000 points and
$\alpha=D=1.2$. The clump is viewed at three different magnifications.
\begin{figure}
\centering
\mbox{{\bf SEE GIF FILE}}
\caption{ A projection of a three dimensional  
Rayleigh-L'evy single clump containing 160000 points generated 
with $\alpha=D=1.2$. The clump is viewed in 3 different magnifications.}
\label{fig:chain}
\end{figure}                    

On scales smaller than the typical length of RL clump in the volume,
the TCH model shares many of the properties of a pure
fractal. However, in the TCH model the mean number density of points
on large scales is well defined.  This allows e.g.  statistical
description of the distribution of points in terms of the two-point
correlation function (Peebles 1980).

\section{The \op\ forest and the dark matter connection}

Absorption spectra are presented in terms of the normalised flux
$F=\exp(-\tau)$ where $\tau$ is the  optical depth  
which is related to the \hi\ density, $\dhi$, along the line-of-sight
 by
\begin{equation}
\tau(w)=  \sigma_{0} \;\frac{c}{H(z)}\;\int_{-\infty}^{\infty} \dhi (x)
{\cal H}[w-x , b(x)]
\dd x,  \\
\label{tau}
\end{equation}
where $\sigma_{0}$ is the effective cross section for resonant line
scattering, $H(z)$ is the Hubble constant at redshift $z$, $x$ is the
real space coordinate (in $\kms$), \calH is the Voigt profile
normalized and $b(x)$ is the
Doppler parameter due to thermal/turbulent broadening. For simplicity 
we have neglected the effect of peculiar velocities. In Cold Dark Matter 
(CDM) models, 
the absorption
features in the \op\ forest are mainly produced by regions of moderate
densities where photoheating is the dominant heating source. 
Because hydrogen in the intergalactic
medium (IGM) is highly ionized (Gunn \& Peterson 1965),
the photoionization equilibrium in the expanding IGM establishes a
tight correlation between neutral and total hydrogen density, $n_{_{\rm H}}$. 
This can
be approximated by a power law $\nhi \propto n_{_{\rm H}}^\beta$,
where 
 $1.56 \la \beta \la 2$
 (Hui \& Gnedin
1997).  Numerical simulations have shown (e.g. Zhang et al. 1995) 
that the total hydrogen density traces
the mass fluctuations on scales larger than the Jeans length. 
So, ionization  equilibrium yields the following relation for 
the  \hi\ density, $\dhi$, 
\begin{equation}
\dhi = \dhhi \left[1+\delta(\vx)\right]^\beta \, , \label{dmhi}
\end{equation}
where
$\delta=\rho/\bar \rho -1$  is the mass density contrast,
 $\bar \rho$ is the mean background mass density, and $\dhhi$
 is the \hi\ density at $\delta=0$.
  The gas density is obtained by smoothing the dark matter  distribution
with the following smoothing window in $k$-space (Bi 1993),
\begin{equation}
W(k)=\frac{1}{1+\left( \frac{k}{ k_J}   \right)^2} , 
\label{smooth}
\end{equation}
where $2\pi/k_J$ is the Jeans scale length which is 
$\sim 1\Mpc$ (comoving) at $z\approx 3$.
We approximate the   
the gas density  field 
by  the dark matter  distribution smoothed with 
the window (\ref{smooth}).

\section{The flux probability distribution function}

As can be seen in Fig 1 a pure RL fractal realization with fractal 
dimension $D<2$ leaves most of space empty.
If the neutral hydrogen  traces faithfully
the dark matter distribution then most of the lines-of-sights to hypothetical
quasars will experience zero optical depth and hence will have
$F=1$. This is very different from what is observed in the real 
universe,  where the mean fluxes $\langle F \rangle$ 
are e.g. $\sim 0.6$ at $z 
\approx 3$.
We  show below that even if we consider the THC model and 
adjust it 
so that the observed value of mean flux is reproduced,
the shape of the flux probability distribution function (PDF) towards a 
single quasar 
significantly differs from the observed PDF.

To make this comparison 
we use two observed high resolution spectra of the QSO 1442+231, from
z=3.6 (the redshift of the QSO)  to z=2.9 (the redshift of Ly$\beta$). 
This redshift range corresponds to a comoving separation
of $250\Mpc$  in a flat $\Omega=1$ universe.
The spectra were observed  by Songalia \& Cowie (1996)
  and Rauch et. al. (1997).
Since the distribution is  a fractal, we can arbitrarily identify the 
length of the box with the redshift interval $z=2.9-3.6$ towards the 
quasar.

We have generated $50$ RL clumps each containing $160000$ points, for
$\alpha=1.2$ and $1.8$, respectively. The clumps were
placed randomly in a cubic box of size unity where $l_0=7 \times 10^{-5}$
and $10^{-3}$ for $\alpha=1.2$ and $1.8$, respectively.  
This values of $l_{0}$ were chosen to be much smaller than 
the mean separation between particles, and to yield clumps of roughly 
equal size in the two values of $\alpha$.
We use  the  clouds-in-cells (CIC) scheme to 
interpolate the  point 
dark matter  distribution in the TCH model to a 
uniform cubic grid. Then we use FFT to convolve 
the gridded density with the window (\ref{smooth}).
In the TCH model, the actual value of the smoothing scale length is not
important as long as it is smaller than the typical size of the clumps
and the mean separation between them.
We use the relations  (\ref{tau})  and (\ref{dmhi}) with $\beta=1.7$
to compute the optical depth and the normalized flux from the smoothed 
dark matter density field in
random  lines-of-sight through the box.
For the comparison 
with the data, we adjust 
 $\dhhi$ in eq. ( \ref{dmhi}) 
so that the mean flux $ \langle F \rangle $
 matches the observed value of $0.65$
over the redshift range $z=2.9-3.6$
for QSO1442+231 (Rauch et al 1997).

\begin{figure}
\centering
\mbox{\psfig{figure=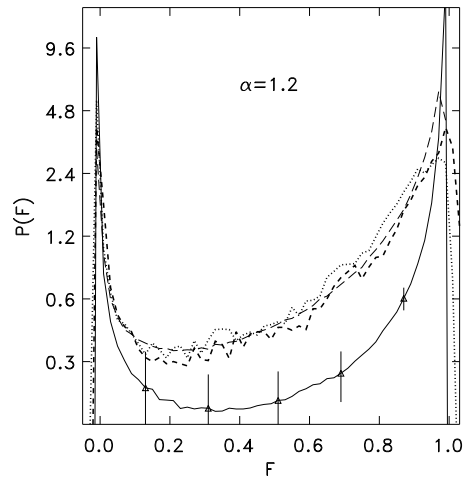,height=7.0cm,width=7.0cm}}
\caption{The PDF of the normalised flux. 
The thick short dashed and dotted 
lines correspond to two observations of the spectrum  
of Q1422. The thin solid line with the error bars is  obtained from 
the TCH model with $\alpha=1.2$. 
For comparison we show as the thin long dashed line the PDF corresponding to 
a lognormal  
 $\Omega=1$ standard CDM  mass distribution normalised  to match the
abundance of rich clusters  in the local universe. }
\label{fig:fpdf}
\end{figure}

\begin{figure}
\centering
\mbox{\psfig{figure=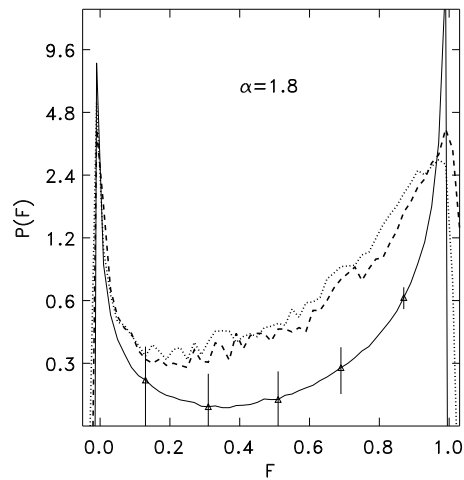,height=7.0cm,width=7.0cm}}
\caption{ 
The same as the previous figure but for $\alpha=1.8$ and without 
the lognormal curve.}
\label{fig:fpdf18}
\end{figure}

 In figures (\ref{fig:fpdf}) and (\ref{fig:fpdf18}) we compare
the models' results with two observed high resolution spectra of the QSO
1422.  
The error bars on the PDF
from the TCH model are rough estimates of the cosmic variance.  In
deriving these error bars we had to fix a physical scale for the box
size. We fix this scale by requiring the mass correlation function at
$z\approx 3$ to be unity at separation $1 \Mpc$ as inferred
by extrapolating the the local galaxy correlation function (e.g.,
Peacock \& Dodds 1994) to $z\approx 3$ in a flat universe $\Omega=1$.
Once we have fixed a physical scale, we can
estimate the cosmic variance for a portion of the quasar spectrum of
of length equal to the one dimensional box size.
For comparison we show the PDF corresponding to 
a lognormal  
 $\Omega=1$ standard CDM  mass distribution normalised  to match the
abundance of rich clusters  in the local universe (Eke et. al. 
1996). The CDM model is clearly a much better fit to the data than the 
TCH model.


\section{Discussion}

The \op\ forest is likely to be a better   tracer of  the mass
density than galaxies and therefore serves  as a better probe
of the density fluctuations at low and high redshifts.
We have  shown that  the RL fractal models (pure as well as
truncated) fail to reproduce the flux   distribution
of the \op\  forest. 
The main reason for this  discrepancy is that these  models 
have low fractal dimension ($D<2$) and  are charactarised  by
large empty  regions (i.e. high  level of `lacunarity').  
 In contrast, models   with  fluctuations  declining  with  
increasing scale-lengths 
and approaching uniformity  on horizon  scale 
(e.g. CDM) are consistent with 
the observed flux distribution.
This result from the \op\ forest   is  in line with 
constraints on the smoothness of the universe on large scales 
from the X-Ray and Microwave Background Radiations, and
from the distribution of radio sources .
Another potential measurement of  the large scale 
smoothness  on scales of $\sim 500 \Mpc$ from the \op\ forest, 
which we have not addressed  here in detail,
can be obtained 
from the  fluctuations in the mean flux in multiple 
lines-of-sight to QSOs.

\section{acknowledgment}
AN thanks the Astronomy department of UC Berkeley, 
the Institute of Astronomy and St. Catharine's College Cambridge 
for the hospitality and support.
OL acknowledges the hospitality of the Technion.
We thank L. Cowie  and M. Rauch for  allowing the use of their data.
We are grateful to M.Davis, M.Haehnelt and M.Rees for stimulating discussions.

\protect
\end{document}